\documentclass[12pt]{article}
\usepackage{hyperref}
\usepackage[english]{babel}
\usepackage{amsmath}
\usepackage{latexsym}
\usepackage{amssymb}
\usepackage[all]{xy}
\usepackage[dvips]{graphicx}
\usepackage{epsfig}
\usepackage{psfrag}

\numberwithin{equation}{section} \numberwithin{table}{section}
\numberwithin{figure}{section}

\begin{document}
%\setlength{\parskip}{10pt}
%\oddsidemargin=0.5cm \evensidemargin=0.5cm \hoffset 0.6cm

%\voffset 0.0cm \hoffset 1.0cm \topmargin 0.0in \evensidemargin 0.0in
%\oddsidemargin 0.0in \textheight 8.6in \textwidth 7.25in
%\parskip 10 pt \pagestyle{plain}

%%%%%%%%%%%%%%%%%%%%%%%%%%%%%%%%%%%%%%%%%%%%%%%%%%%%%%%%%%%%%%%%%%%%%%%%
% title page
%%%%%%%%%%%%%%%%%%%%%%%%%%%%%%%%%%%%%%%%%%%%%%%%%%%%%%%%%%%%%%%%%%%
\begin{titlepage}
  \begin{flushright}
 {\small CQUeST-2009-0265}
  \end{flushright}

  \begin{center}

    \vspace{20mm}

    {\LARGE \bf A Note on Black Holes in Asymptotically Lifshitz Spacetime}

    \vspace{10mm}

    %Shan Bai$^{\dag}$,
    Da-Wei Pang$^{\dag}$

    \vspace{5mm}
    %{\small \sl $\dag$ Key Laboratory of Frontiers in Theoretical
%    Physics£¬\\
%    Institute of Theoretical Physics£¬Chinese Academy of Sciences}\\
%    {\small \sl P.O.Box 2735, Beijing 100190, China} \\
    {\small \sl $\dag$ Center for Quantum Spacetime, Sogang University}\\
    {\small \sl Seoul 121-742, Korea\\}
    {\small \tt pangdw`at'sogang.ac.kr}
    \vspace{10mm}

  \end{center}

\begin{abstract}
\baselineskip=18pt We investigate several aspects of exact black
hole solutions in asymptotically Lifshitz spacetime, which were
proposed in 0812.0530. Firstly, we calculate the tidal forces and
find that in the near horizon region of such black hole backgrounds,
the tidal forces diverge in the near extremal limit. Secondly, we
evaluate the Wilson loops in both extremal and finite temperature
cases. Finally, we obtain the corresponding shear viscosity and
square of the sound speed and find that the ratio of shear viscosity
to entropy density takes the universal value $1/4\pi$ in arbitrary
dimensions while the square of the speed of sound saturates the
conjectured bound $1/3$ in five dimensions.
\end{abstract}
\setcounter{page}{0}
\end{titlepage}

\pagestyle{plain} \baselineskip=19pt

\tableofcontents

%%%%%%%%%%%%%%%%%%%%%%%%%%%%%%%%%%%%%%%%%%%%%%%%%%%%%%%%%%%%%%%%%%%%%%%%%%%%%%%%%%%%%%%%%%%%%%%%%
\section{Introduction}
%%%%%%%%%%%%%%%%%%%%%%%%%%%%%%%%%%%%%%%%%%%%%%%%%%%%%%%%%%%%%%%%%%%%%%%%%%%%%%%%%%%%%%%%%%%%%%%%%
By now, the AdS/CFT
correspondence~\cite{Maldacena:1997re,Gubser:1998bc, Witten:1998qj}
is the unique approach which relates strongly coupled field theories
to weakly coupled gravity theory. It has been extensively
investigated in the past decade and its validity has been widely
recognized in the theoretical high-energy physics community.
Recently there has been enormous progress on the application of
AdS/CFT correspondence, or even the more general gauge/gravity
correspondence to physical systems in the real world, such as
AdS/QCD and holographic methods for condensed matter physics. Two
nice reviews are given by~\cite{Son:2007vk, Hartnoll:2009sz}.

It is well known that certain questions which are difficult to deal
with in the field theory side, become more transparent and more
tractable in the gravity side via the AdS/CFT correspondence. In
condensed matter physics there are many strongly coupled systems, so
it is widely hoped that the AdS/CFT correspondence can provide some
useful tools for studying condensed matter physics. Recently
interesting gravity models dual to various condensed matter systems
have been proposed~\cite{Hartnoll:2008vx}-\cite{Adams:2008wt}.

Special attention has been paid to gravity duals of Lifshitz-like
fixed points, which is initially proposed in~\cite{Kachru:2008yh}.
One can find critical phenomena with unconventional scaling behavior
in many condensed matter systems
\begin{equation}
\label{1eq1} t~\rightarrow~\lambda^{z}t,~~~~~{\bf
x}~\rightarrow~\lambda{\bf x},
\end{equation}
where $z\neq1$. A toy model realizing this scaling behavior with
$z=2$ is the so-called Lifshitz field theory,
\begin{equation}
\label{1eq2} \mathcal{L}=\int
d^{2}xdt((\partial_{t}\phi)^2-\kappa(\nabla^{2}\phi)^2).
\end{equation}
The corresponding gravity dual takes the following
form~\cite{Kachru:2008yh}
\begin{equation}
\label{1eq3}
ds^{2}=L^{2}(-r^{2z}dt^{2}+\frac{dr^{2}}{r^{2}}+r^{2}d{\bf x}^2),
\end{equation}
where $d{\bf x}^2=dx^{2}_{1}+\cdots dx^{2}_{d}$. This metric
exhibits the following scale invariance
\begin{equation}
t~\rightarrow~\lambda^{z}t,~~~~r~\rightarrow~\frac{r}{\lambda},~~~~{\bf
x}~\rightarrow~\lambda{\bf x}.
\end{equation}
Note that when $z=1$, it turns out to be the usual $AdS_{d+2}$
spacetime. In four-dimensional spacetime, the corresponding action
is a gravity theory with negative cosmological constant, coupled
with abelian gauge fields $A_{(1)}, B_{(2)}$
\begin{equation}
\label{1eq5} S=\int d^{4}x\sqrt{-g}(R-2\Lambda)-\frac{1}{2}\int(\ast
F_{(2)}\wedge F_{(2)}+\ast H_{(3)}\wedge H_{(3)})-c\int
B_{(2)}\wedge F_{(2)},
\end{equation}
where $F_{(2)}=dA_{(1)}, H_{(3)}=dB_{(2)}$ and the cosmological
constant $\Lambda=-5/L^{2}$.

There are also several generalizations along a similar way. Various
anisotropic gravity solutions in general spacetime dimensions with
different scaling behavior were discussed in~\cite{SSPal}. The
aspects of holography in general anisotropic, non-relativistic
backgrounds were investigated extensively in~\cite{Taylor:2008tg}.
The geometry of Lifshitz spacetime was studied
in~\cite{Blau:2009gd}. Furthermore, the embedding of such
anisotropic gravity background with somewhat different scaling
behavior into string theory was realized quite recently in~
\cite{Azeyanagi:2009pr}, where the corresponding black brane
configurations were also obtained.

However, although we can study the zero-temperature Lifshitz
spacetime, it is difficult to obtain exact black hole solutions in
the context of the original action~(\ref{1eq5}). The black hole in
asymptotically Lifshitz spacetime was constructed
in~\cite{Danielsson:2009gi} using numerical methods, while Lifshitz
topological black holes were obtained in~\cite{Mann:2009yx}. It
should be pointed out that an exact solution of topological Lifshitz
black holes was obtained in~\cite{Mann:2009yx} for certain
particular case.

When discussing the aspects of holography in more general
anisotropic, non-relativistic backgrounds in~\cite{Taylor:2008tg},
an exact solution with finite temperature was obtained by making use
of a different action. It can be seen that by performing some
coordinate transformation, the black hole solution is asymptotically
Lifshitz-like. In this note we discuss several aspects of this exact
solution. In section 2 we rewrite the black hole solution in a more
transparent way. Then in section 3 we calculate the tidal forces and
it turns out that such tidal forces become divergent in the near
horizon region, while the horizon area remains large. In this sense,
this type of Lifshitz black hole is
``naked''~\cite{Horowitz:1997uc}. In the next section we evaluate
the Wilson loops in this asymptotically Lifshitz black hole
background, the results agree with previous examples in the extremal
limit and the finite temperature cases are calculated numerically.
We discuss the hydrodynamic properties in section 5. It can be shown
that the ratio of shear viscosity to entropy density is $1/4\pi$ in
arbitrary dimensions and the square of the speed of sound is $1/3$
in five dimensional spacetime, both of which saturate the well-known
bounds. A summary and discussion on further directions are given in
the final section.
%%%%%%%%%%%%%%%%%%%%%%%%%%%%%%%%%%%%%%%%%%%%%%%%%%%%%%%%%%%%%%%%%%%%%%%%%%%%%%%%%%%%%%%%%%%%%%%%%%%%
\section{The black hole solution}
%%%%%%%%%%%%%%%%%%%%%%%%%%%%%%%%%%%%%%%%%%%%%%%%%%%%%%%%%%%%%%%%%%%%%%%%%%%%%%%%%%%%%%%%%%%%%%%%%%%%
In this section we review the asymptotic Lifshitz solutions proposed
in~\cite{Taylor:2008tg}, including the extremal solution and black
hole solution. We will rewrite the solutions in a more transparent
way.

Consider the following action in $(d+2)$-dimensional spacetime
\begin{equation}
S=\frac{1}{16\pi G_{d+2}}\int
d^{d+2}x\sqrt{-g}[R-2\Lambda-\frac{1}{2}\partial_{\mu}\phi\partial^{\mu}\phi
-\frac{1}{4}e^{\lambda\phi}F_{\mu\nu}F^{\mu\nu}],
\end{equation}
where $\Lambda$ is the cosmological constant and the matter fields
are a massless scalar and an abelian gauge field. The equations of
motion can be written as follows:
\begin{equation}
\label{2eq2}
\partial_{\mu}(\sqrt{-g}e^{\lambda\phi}F^{\mu\nu})=0,
\end{equation}
\begin{equation}
\partial_{\mu}(\sqrt{-g}\partial^{\mu}\phi)-\frac{\lambda}{4}\sqrt{-g}e^{\lambda\phi}F_{\mu\nu}F^{\mu\nu}=0,
\end{equation}
\begin{equation}
\label{2eq4} R_{\mu\nu}=\frac{2}{d}\Lambda
g_{\mu\nu}+\frac{1}{2}\partial_{\mu}\phi\partial_{\nu}\phi+\frac{1}{2}e^{\lambda\phi}F_{\mu\rho}{F_{\nu}}^{\rho}
-\frac{1}{4d}g_{\mu\nu}e^{\lambda\phi}F_{\mu\nu}F^{\mu\nu}.
\end{equation}
We make the following ansatz for the metric
\begin{equation}
ds^{2}=L^{2}[-r^{2z}f(r)dt^{2}+\frac{dr^{2}}{r^{2}f(r)}+r^{2}\sum\limits^{d}_{i=1}dx^{2}_{i}],
\end{equation}
where $z\geq1$ and the only non-vanishing component of the field
strength is $F_{rt}$.

We can obtain the following expression for $F_{rt}$ by
solving~(\ref{2eq2})
\begin{equation}
F_{rt}=qe^{-\lambda\phi}r^{z-d-1},
\end{equation}
where $q$ is a constant which can be related to the charge of the
black hole. Furthermore, solving the $tt$ and $rr$ components
of~(\ref{2eq4})we can arrive at
\begin{equation}
\label{2eq7}
\partial_{r}\phi\partial_{r}\phi=\frac{2(z-1)d}{r^{2}}.
\end{equation}
When $z=1$, it can be easily seen that the solution is
$\phi=\phi_{0}={\rm const}$. The full solution can be obtained by
solving the remaining equations of motion
\begin{eqnarray}
&
&ds^{2}=L^{2}[-r^{2}dt^{2}+\frac{dr^{2}}{r^{2}}+r^{2}\sum\limits^{d}_{i=1}dx^{2}_{i}],\nonumber\\
& &\phi={\rm const},~~~F_{rt}=0,~~~\Lambda=-\frac{d(d+1)}{2L^{2}}.
\end{eqnarray}
It is simply the AdS solution in Poincar\'{e} coordinates. It also
admits black hole solution with $f(r)=1-r^{d+1}_{+}/r^{d+1}$ and
other fields remaining the same as the AdS solution.

When $z\neq1$, from~(\ref{2eq7}) we can obtain
\begin{equation}
\phi=\pm\sqrt{2(z-1)d}\log r,
\end{equation}
where we have taken the integration constant to be zero without loss
of generality. Similarly, we can summarize the extremal solution as
follows
\begin{eqnarray}
&
&ds^{2}=L^{2}(-r^{2z}dt^{2}+\frac{dr^{2}}{r^{2}}+r^{2}\sum\limits^{d}_{i=1}dx^{2}_{i}),\nonumber\\
&
&F_{rt}=qe^{-\lambda\phi}r^{z-d-1},~~~e^{\lambda\phi}=r^{\lambda\sqrt{2(z-1)d}},\nonumber\\
& &\lambda^{2}=\frac{2d}{z-1},~~~q^{2}=2L^{2}(z-1)(z+d),\nonumber\\
& &\Lambda=-\frac{(z+d-1)(z+d)}{2L^{2}}.
\end{eqnarray}
It is just the Lifshitz spacetime with non-trivial dilaton and gauge
fields. It should be pointed out that the finite temperature
generalization
\begin{equation}
\label{2eq11}
ds^{2}=L^{2}(-r^{2z}f(r)dt^{2}+\frac{dr^{2}}{r^{2}f(r)}+r^{2}\sum\limits^{d}_{i=1}dx^{2}_{i}),~~~
f(r)=1-\frac{r^{z+d}_{+}}{r^{z+d}},
\end{equation}
is also a solution to the equations of motion with the same field
configuration. Thus the finite temperature solution is an
asymptotically Lifshitz black hole.

Now let us focus on the asymptotically Lifshitz black hole solution.
The temperature is
\begin{equation}
\label{2eq12} T_{H}=\frac{(z+d)r^{z}_{+}}{4\pi},
\end{equation}
and the black hole entropy is
\begin{equation}
\label{2eq13} S_{BH}=\frac{V_{d}}{4G_{d+2}}L^{d}r^{d}_{+},
\end{equation}
where $V_{d}$ denotes the volume of the $d$ dimensional spatial
coordinates. One can rewrite the entropy as a function of
temperature
\begin{equation}
S_{BH}=\frac{V_{d}L^{d}}{4G_{d+2}}(\frac{4\pi}{z+d})^{\frac{d}{z}}T^{\frac{d}{z}},
\end{equation}
which exhibits the expected behavior of an anisotropic scale
invariant theory.

The thermodynamic quantities can be obtained via the Euclidean path
integral method, which were calculated explicitly
in~\cite{Taylor:2008tg}. Here we shall not dwell on the details but
only list some useful results. Consider the following Euclidean
action
\begin{equation}
I_{E}=-\frac{1}{16\pi G_{d+2}}\int
d^{d+2}x\sqrt{-g}[R-2\Lambda-\frac{1}{2}\partial_{\mu}\phi\partial^{\mu}\phi
-\frac{1}{4}e^{\lambda\phi}F_{\mu\nu}F^{\mu\nu}]-\frac{1}{8\pi
G_{d+2}}\int d^{d+1}x\sqrt{h}K,
\end{equation}
where the second term is the Gibbons-Hawking boundary term. After
substituting the background configuration, the Euclidean action
turns out to be
\begin{equation}
\label{2eq16} I_{E}=-\frac{r^{z+d}_{+}L^{d}V_{d}\beta_{H}}{16\pi
G_{d+2}},
\end{equation}
where $\beta_{H}=1/T_{H}$. We can calculate the other thermodynamic
quantities in a standard way as soon as we obtain the Euclidean
action. For example, the mass of the black hole is
\begin{equation}
\label{2eq17} M=\frac{r^{z+d}_{+}dL^{d}V_{d}}{16\pi G_{d+2}},
\end{equation}
and the charge of the black hole is given by
\begin{equation}
\label{2eq18} Q=\frac{1}{32\pi G_{d+2}}\int e^{\lambda\phi}(\ast F)=
\frac{qL^{d}V_{d}}{32\pi G_{d+2}}. \end{equation}

From~(\ref{2eq16}) we can see that there is no interesting phase
structure for this asymptotically Lifshitz black hole, as the
Euclidean action is always negative. This can also be seen from the
heat capacity
\begin{equation}
C=\frac{dM}{dT}=\frac{\partial M/\partial r_{+}}{\partial T/\partial
r_{+}}.
\end{equation}
Using~(\ref{2eq12}) and~(\ref{2eq17}), we can obtain
\begin{equation}
C=\frac{dV_{d}r^{d}_{+}}{4zG_{d+2}},
\end{equation}
which shows that the black hole is always thermodynamically stable.
%%%%%%%%%%%%%%%%%%%%%%%%%%%%%%%%%%%%%%%%%%%%%%%%%%%%%%%%%%%%%%%%%%%%%%%%%%%%%%%%%%%%%%%%%%%%%%%%%%%%
\section{Tidal forces}
%%%%%%%%%%%%%%%%%%%%%%%%%%%%%%%%%%%%%%%%%%%%%%%%%%%%%%%%%%%%%%%%%%%%%%%%%%%%%%%%%%%%%%%%%%%%%%%%%%%%
In this section we will calculate the tidal forces of the Lifshitz
black hole, following~\cite{Horowitz:1997uc}. It has been shown that
there exist a class of black holes whose horizon area is large and
all curvature invariants are small near the horizon, while any
object falling in experiences large tidal forces outside the
horizon. As the region of large tidal forces is visible to distant
observers, such black holes are called ``naked''.

Recall the metric
\begin{equation}
ds^{2}=L^{2}(-r^{2z}f(r)dt^{2}+\frac{dr^{2}}{r^{2}f(r)}+r^{2}\sum\limits^{d}_{i=1}dx^{2}_{i}),~~~
f(r)=1-\frac{r^{z+d}_{+}}{r^{z+d}},
\end{equation}
and the vielbein in the static frame is given as
\begin{equation}
(e_{0})_{\mu}=-Lr^{z}f(r)^{1/2}\partial_{\mu}t,~~~(e_{1})_{\mu}=Lr^{-1}f(r)^{-1/2}\partial_{\mu}r,
~~~(e_{i})_{\mu}=Lr\partial_{\mu}x_{i}.
\end{equation}
Consider timelike geodesics in the above background, with proper
time $\tau$ and tangent vector $u^{\mu}=dx^{\mu}/d\tau$. The
constants of motion can be written as follows
\begin{equation}
E=L^{2}r^{2z}f(r)\dot{t},~~~p_{i}=L^{2}r^{2}\dot{x}_{i},
\end{equation}
where an overdot denotes $d/d\tau$. For simplicity, we just consider
radial geodesics, i.e. $p_{i}=0$. We can arrive at the following
expression due to the normalization condition $u^{\mu}u_{\mu}=-1$
\begin{equation}
\dot{r}^{2}=\frac{E^{2}}{L^{4}r^{2z-2}}-\frac{r^{2}}{L^{2}}f(r).
\end{equation}
The parallel-propagated orthonormal frame
$(e_{0^{\prime}})_{\mu}=u_{\mu}$ can be obtained by a boost of the
original static frame
\begin{eqnarray}
(e_{0^{\prime}})_{\mu}&=&u_{\mu}=-E\partial_{\mu}t+\frac{\dot{r}L^{2}}{r^{2}f(r)}\partial_{\mu}r\nonumber\\
&\equiv&\cosh\alpha(e_{0})_{\mu}+\sinh\alpha(e_{1})_{\mu}\nonumber\\
(e_{1^{\prime}})_{\mu}&=&\sinh\alpha(e_{0})_{\mu}+\cosh\alpha(e_{1})_{\mu},
\end{eqnarray}
where $\cosh\alpha=E[L^{2}r^{2z}f(r)]^{-1/2}$ and the other
components remain invariant. It can be seen that the boost parameter
$\alpha$ diverges at the horizon.

The components of the Riemann curvature in the boosted frame can be
calculated by working out the components in the static frame first
and then performing some transformations. However, there is another
simple route to calculate such quantities which has a more direct
physical meaning~\cite{Horowitz:1997uc}. We will calculate
$R_{0^{\prime}k0^{\prime}k}$ which correspond to tidal forces in the
transverse directions. Consider a class of radial infalling
geodesics whose tangent vector is $u^{\mu}$ and the deviation
vectors are $\eta^{i}=\partial/\partial x_{i}$, we have
\begin{equation}
u^{\nu}\nabla_{\nu}\eta^{\sigma}=u^{\nu}\Gamma^{\sigma}_{\nu\rho}\eta^{\rho}=\frac{\dot{H}}{H}\eta^{\sigma},
\end{equation}
where $H=Lr$. Thus the geodesic deviation equation gives
\begin{equation}
{R_{\mu\nu\rho}}^{\sigma}u^{\mu}\eta^{\nu}u^{\rho}=-u^{\mu}\nabla_{\mu}(u^{\nu}\nabla_{\nu}\eta^{\sigma})
=-\frac{\ddot{H}}{H}\eta^{\sigma}.
\end{equation}
Therefore
\begin{eqnarray}
R_{0^{\prime}i0^{\prime}i}&=&{R_{\mu\nu\rho}}^{\sigma}u^{\mu}(e_{i})^{\nu}u^{\rho}(e_{i})_{\sigma}
=-\frac{\ddot{H}}{H}\nonumber\\
&=&\frac{(z-1)E^{2}}{L^{4}r^{2z}}+\frac{1}{L^{2}}[1+\frac{(z+d-2)r^{z+d}_{+}}{2r^{z+d}}].
\end{eqnarray}

The enhancement of the curvature in the geodesic frame leads to the
term proportional to $E^{2}$ in the above expression. It can be seen
that if we take the conserved quantity $E$ to be very large, the
tidal force can be made arbitrarily large. Conversely, we can also
make the tidal force very small. Thus in order to avoid such
ambiguities, we assume that the conserved quantity $E$ is chosen to
be order one. It is sufficient to keep the term proportional to
$E^{2}$ only, as such term represents the difference between the
static frame and the boosted frame. Then the tidal force in the near
horizon region is given by
\begin{equation}
R_{0^{\prime}i0^{\prime}i}=\frac{(z-1)E^{2}}{L^{4}r^{2z}_{+}}.
\end{equation}
It can be easily seen that the tidal force vanishes in $z=1$ case
then we will consider two different near-extremal limits with the
assumption $z>1$ in the following.
\begin{itemize}
\item $r_{+}<<1$ with fixed mass.\\
Recalling~(\ref{2eq17}), we can arrive at the following equation
\begin{equation}
M {\rm fixed}~~~\rightarrow~~~L\sim r^{-(z+d)/d}.
\end{equation}
Then the tidal force becomes
\begin{equation}
R_{0^{\prime}i0^{\prime}i}=(z-1)E^{2}r^{4+4z/d-2z}_{+}.
\end{equation}
The horizon area satisfies
\begin{equation}
A\propto L^{d}r^{d}_{+}\sim r^{-z}_{+}.
\end{equation}
So $r_{+}<<1$ makes the horizon area large. However, when $d=2$, the
tidal force is $(z-1)E^{2}r^{4}_{+}$, that is, the tidal force also
turns out to be very small. In this limit the black holes are not
``naked''. When $d>2$, the requirement that the tidal force is large
gives
\begin{equation}
4+4\frac{z}{d}-2z<0~~~\rightarrow~~~z>\frac{4d}{2d-4}=2+\frac{8}{2d-4}.
\end{equation}
It can be seen that the $z=2$ case can never lead to ``naked'' black
holes.
\item $r_{+}<<1$ without fixing the mass.\\
Here the requirement that the horizon area is large gives
\begin{equation}
A\propto L^{d}r^{d}_{+}>>1~~~\rightarrow~~~Lr_{+}>>1.
\end{equation}
The tidal force turns out to be
\begin{equation}
R_{0^{\prime}i0^{\prime}i}\propto\frac{1}{(Lr_{+})^{4}r^{2z-4}_{+}}.
\end{equation}
Thus it is possible to have ``naked'' black holes only in the $z>2$
cases.
\end{itemize}
%%%%%%%%%%%%%%%%%%%%%%%%%%%%%%%%%%%%%%%%%%%%%%%%%%%%%%%%%%%%%%%%%%%%%%%%%%%%%%%%%%%%%%%%%%%%%%%%%%%%
\section{Wilson loops}
%%%%%%%%%%%%%%%%%%%%%%%%%%%%%%%%%%%%%%%%%%%%%%%%%%%%%%%%%%%%%%%%%%%%%%%%%%%%%%%%%%%%%%%%%%%%%%%%%%%%
In this section we study Wilson loops for asymptotically Lifshitz
black holes. The Wilson loops describe the behavior of quarks by
hanging strings from the boundary where the quarks locate at the
ends of the strings. Although it is quite difficult to embed
Lifshitz spacetime into string theory, the calculations presented
here can provide some qualitative information. Consider rectangular
Wilson loops in Euclidean spacetime, the dynamics is described by
the Nambu-Goto action
\begin{equation}
S=-\int d\sigma\sqrt{{\rm
det}h_{ab}},~~~h_{ab}=g_{\mu\nu}\partial_{a}X^{\mu}(\tau,\sigma)\partial_{b}X^{\nu}(\tau,\sigma),
\end{equation}
where $X^{\mu}(\tau,\sigma)$ denote the string coordinates and
$\tau, \sigma$ parametrize the string worldsheet. In the following
we will focus on five-dimensional asymptotic Lifshitz black holes,
whose metric is
\begin{equation}
ds^{2}=L^{2}[-r^{2z}f(r)dt^{2}+\frac{dr^{2}}{r^{2}f(r)}+r^{2}(dx^{2}_{1}+dx^{2}_{2}+dx^{2}_{3})].
f(r)=1-\frac{r^{z+3}_{+}}{r^{z+3}}.
\end{equation}
Then we can obtain the equations of motion
\begin{eqnarray}
&
&(\frac{r^{2z+2}f(r)x^{\prime}}{\sqrt{r^{2z-2}r^{\prime2}+r^{2z+2}f(r)x^{\prime2}}})^{\prime}=0,
\nonumber\\
&
&(\frac{r^{2z-2}r^{\prime}}{\sqrt{r^{2z-2}r^{\prime2}+r^{2z+2}f(r)x^{\prime2}}})^{\prime}=\frac{1}{2}
\frac{(2z-2)r^{2z-3}r^{\prime2}+\partial_{r}(r^{2z+2}f(r))x^{\prime2}}
{\sqrt{r^{2z-2}r^{\prime2}+r^{2z+2}f(r)x^{\prime2}}},
\end{eqnarray}
where the prime stands for derivative with respect to $\sigma$.

One possible static configuration is a pair of straight macroscopic
strings which are stretched between $r=\infty$ and $r=r_{+}$. The
corresponding total energy is
\begin{equation}
E_{0}=2L^{2}\int^{\infty}_{r_{+}}r^{z-1}dr.
\end{equation}
The other possible configuration is a macroscopic U-shape string
whose each end is connected to the quark and anti-quark at the
boundary. In the static gauge $\sigma=x$, the equations of motion
turn out to be
\begin{eqnarray}
&
&(\frac{r^{2z+2}f(r)}{\sqrt{r^{2z-2}r^{\prime2}+r^{2z+2}f(r)}})^{\prime}=0,
\nonumber\\
&
&(\frac{r^{2z-2}r^{\prime}}{\sqrt{r^{2z-2}r^{\prime2}+r^{2z+2}f(r)}})^{\prime}=\frac{1}{2}
\frac{(2z-2)r^{2z-3}r^{\prime2}+\partial_{r}(r^{2z+2}f(r))}
{\sqrt{r^{2z-2}r^{\prime2}+r^{2z+2}f(r)}}.
\end{eqnarray}
 We can
arrive at the following result by extremizing the action
\begin{equation}
\frac{f(r)r^{2z+2}}{\sqrt{r^{2z-2}r^{\prime2}+f(r)r^{2z+2}}}={\rm
const}=f^{1/2}_{\rm min}r^{z+1}_{\rm min},
\end{equation}
where $r_{\rm min}$ is $r$ coordinate of the string tip which is the
closest to the horizon and $f_{\rm min}\equiv f(r)|_{r=r_{\rm
min}}$. Note that $\partial r/\partial x=0$ at $r=r_{\rm min}$. From
the above expression we can rewrite $x$ as a function of $r$
\begin{equation}
x=\int^{r}_{r_{\rm
min}}dr\frac{1}{r^{2}f(r)^{1/2}\sqrt{(\frac{r}{r_{\rm
min}})^{2z+2}(\frac{f}{f_{\rm min}})-1}}.
\end{equation}
Thus the boundary distance between the endpoints of the string is
given by
\begin{equation}
\ell=2\int^{\infty}_{r_{\rm
min}}dr\frac{1}{r^{2}f(r)^{1/2}\sqrt{(\frac{r}{r_{\rm
min}})^{2z+2}(\frac{f}{f_{\rm min}})-1}}.
\end{equation}
The total energy of the U-shape string with inter-quark separation
$\ell$ is
\begin{eqnarray}
E&=&L^{2}\int^{\frac{\ell}{2}}_{-\frac{\ell}{2}}dx\sqrt{r^{2z-2}
r^{\prime2}+r^{2z+2}f(r)}\nonumber\\
&=&2L^{2}\int dr\frac{r^{4}f(r)^{1/2}}{\sqrt{r^{6}f(r)-r^{6}_{\rm
min}f_{\rm min}}}.
\end{eqnarray}
Finally, the heavy quark potential is given by
\begin{equation}
V=E-E_{0},
\end{equation}
where we have subtracted the contribution of two straight strings.

In the extremal background, i.e. $r_{+}=0$, it can be seen that
these results agree with those given in~\cite{Danielsson:2009gi} and
these expressions reduce to those of~\cite{Rey:1998ik}
and~\cite{Maldacena:1998im} when $z=1$. In the finite temperature
case~\cite{Rey:1998bq}, the integration can be worked out
analytically by making use of the elliptic integral. Unfortunately,
here we cannot obtain analytical results thus we have to evaluate
the integrals numerically. For simplicity we just consider the
five-dimensional case with $z=2$, that is, $f=1-r^{5}_{+}/r^{5}$.
The expressions for the boundary distance $\ell$ and the potential
energy can be rewritten as follows
\begin{equation}
\ell=\frac{2}{r_{+}}\sqrt{a^{6}-a}\int^{\infty}_{a}\frac{\sqrt{y}}
{\sqrt{(y^{5}-1)[(y^{6}-a^{6})-(y-a)]}}dy,
\end{equation}
and
\begin{equation}
V=2r^{2}_{+}L^{2}[\int^{\infty}_{a}dy(\frac{y^{3/2}\sqrt{y^{5}-1}}{\sqrt{(y^{6}-a^{6})-(y-a)}}-y)
-\frac{1}{2}(a^{2}-1)],
\end{equation}
where we have introduced $y\equiv r/r_{+}$ and $a\equiv r_{\rm
min}/r_{+}$. Note that the parameter $a$ should be larger than 1,
that is, the string always stays outside the horizon.

\begin{figure}
\centering
\includegraphics[width=10cm]{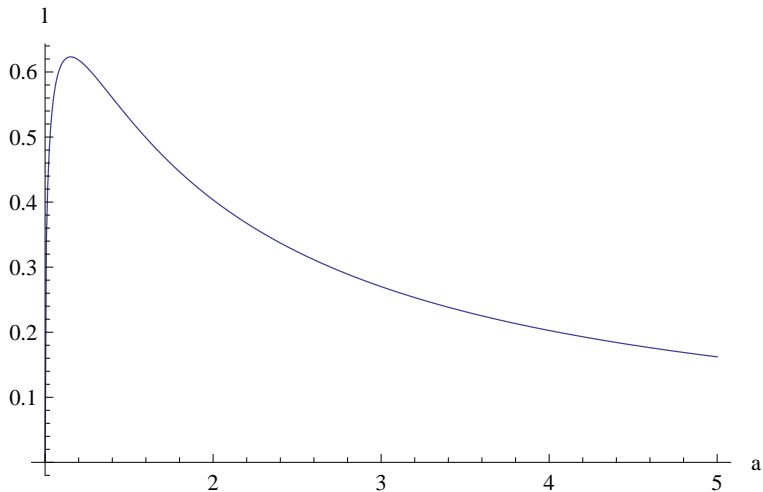}
\caption{The boundary distance between the endpoints of a string
$\ell$ as a function of $a$, with $r_{+}=1$.} \label{fig:1}
\end{figure}

\begin{figure}
\centering
\includegraphics[width=10cm]{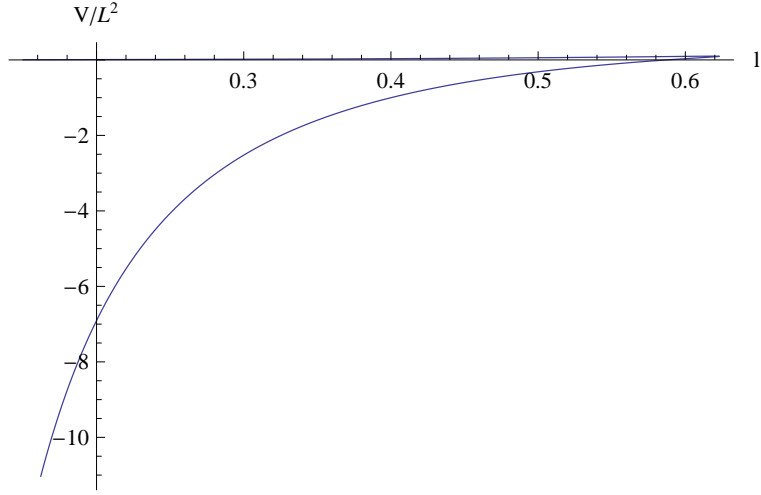}
\caption{The potential energy as a function of the boundary distance
$\ell$, with $r_{+}=1$.} \label{fig:2}
\end{figure}

The boundary distance between the endpoints of a string $\ell$ as a
function of $a$ is shown in Fig.~\ref{fig:1}, while the potential
energy as a function of the boundary distance $\ell$ is shown in
Fig.~\ref{fig:2}. Compared to the results in~\cite{Rey:1998bq}, it
can be seen that these functions exhibit similar behavior. The
boundary distance between the endpoints of a string has a maximum
value $\ell_{\rm max}$. For a fixed $\ell<\ell_{\rm max}$, there are
two possible U-shape string configurations at two different values
of $a$. The energy of the U-shape string is plotted in
Fig.~\ref{fig:2}. The configuration with smaller $a$ has a nearly
zero potential energy and the configuration with larger $a$ has
lower energy. The potential crosses zero at $\ell=\ell_{\ast}$. The
pair of straight strings has lower energy than the U-shape string
configuration once $\ell>\ell_{\ast}$.

%%%%%%%%%%%%%%%%%%%%%%%%%%%%%%%%%%%%%%%%%%%%%%
%%%%%%%%%%%%%%%%%%%%%%%%%%%%%%%%%%%%%%%%%%%%%%%%%%

%%%%%%%%%%%%%%%%%%%%%%%%%%%%%%%%%%%%%%%%%%%%%%
%%%%%%%%%%%%%%%%%%%%%%%%%%%%%%%%%%%%%%%%%%%%%%%%%%

%%%%%%%%%%%%%%%%%%%%%%%%%%%%%%%%%%%%%%%%%%%%%%%%%%%%%%%%%%%%%%%%%%%%%%%%%%%%%%%%%%%%%%%%%%%%%%%%%%%%
\section{Hydrodynamics}
%%%%%%%%%%%%%%%%%%%%%%%%%%%%%%%%%%%%%%%%%%%%%%%%%%%%%%%%%%%%%%%%%%%%%%%%%%%%%%%%%%%%%%%%%%%%%%%%%%%%
In this section we discuss the hydrodynamic properties of such
asymptotically Lifshitz black holes, including the shear viscosity
and the speed of sound. We will see that the ratio of shear
viscosity to entropy density is $1/4\pi$ in arbitrary dimensions,
which saturates the well known KSS bound~\cite{Kovtun:2004de}, while
the square of the speed of sound is $1/d$. It should be pointed out
that in five-dimensional case, the square of the speed of sound is
$1/3$, which also saturates the bound proposed very
recently~\cite{Hohler:2009tv},~\cite{Cherman:2009tw}.

Firstly let us focus on the shear viscosity $\eta$. We will apply
the Kubo formula
\begin{equation}
\eta=-\lim\limits_{\omega\rightarrow0}\frac{1}{\omega}{\rm
Im}G^{R}(\omega,\vec{k}=0),
\end{equation}
where $G^{R}$ is the retarded two-point function of the scalar mode
of the stress tensor
\begin{equation}
G^{R}(\omega,\vec{k}=0)=-i\int d^{d}xdte^{i\omega
t}\theta(t)<[T_{xy}(t,\vec{x}),T_{xy}(0,0)]>.
\end{equation}
Following the prescription proposed in~\cite{Kovtun:2003wp}, the
linearized Einstein equation for $\phi\equiv h^{x}_{y}(r)e^{-i\omega
t}$ is the scalar wave function in the same background, due to the
$SO(2)$ symmetry of rotations in the $xy-$plane.

Recalling the black hole solution~(\ref{2eq11}), we make coordinate
transformation $u^{2}=r^{z+d}_{+}/r^{z+d}$ for convenience. Then the
black hole metric turns out to be
\begin{equation}
ds^{2}=L^{2}[-(\frac{r^{z+d}_{+}}{u^{2}})^{\frac{2z}{z+d}}f(u)dt^{2}+\frac{4}{(z+d)^{2}u^{2}f(u)}
du^{2}+(\frac{r^{z+d}_{+}}{u^{2}})^{\frac{2}{z+d}}\sum\limits^{d}_{i=1}dx^{2}_{i}],~~~f(u)=1-u^{2}.
\end{equation}
Assuming $\Phi(\omega, u)=\phi(u)e^{-i\omega t}$, the scalar wave
equation
\begin{equation}
\Box\Phi=\frac{1}{\sqrt{-g}}\partial_{\mu}(\sqrt{-g}g^{uu}\partial_{u}\phi)+g^{tt}\partial_{t}\partial_{t}\phi
\end{equation}
gives
\begin{equation}
\label{5eq5}
u^{3}f(u)\partial_{u}(\frac{f(u)}{u}\partial_{u}\phi)+u^{\frac{4z}{z+d}}\beta^{2}\omega^{2}\phi=0,
\end{equation}
where $\beta^{-1}\equiv\frac{1}{2}r^{z}_{+}(z+d)$. Following the
standard procedure, we set $\phi_{k}=(1-u)^{\alpha}$ and require
that the most singular terms at $u=1$ cancel as well as the incoming
wave boundary condition. These requirements finally fix
\begin{equation}
\alpha=-\frac{i}{2}\beta\omega.
\end{equation}
Next, we choose
$\phi_{k}(u)=(1-u)^{-\frac{i}{2}\beta\omega}(1+\frac{i}{2}\beta\omega
F_{1}(u))$, then substitute this expression back to the scalar wave
function~(\ref{5eq5}). Note that in order to calculate the shear
viscosity, we just need the perturbation up to the first order of
$\omega$. Furthermore, $F_{1}(u)$ should be zero at $u=1$. It can be
easily obtained that
\begin{equation}
\phi_{k}(u)=(1-u)^{-\frac{i}{2}\beta\omega}(1-\frac{i}{2}\beta\omega\ln\frac{1+u}{2}).
\end{equation}

Finally combining the flux factor
\begin{equation}
\mathcal{F}=K\sqrt{-g}g^{uu}\phi_{-k}(u)\partial_{u}\phi_{k}(u),
\end{equation}
where $K=-\frac{1}{32\pi G_{d+2}}$ is the coupling constant and the
retarded green function
\begin{equation}
G^{R}(\omega,\vec{k}=0)=-2\mathcal{F}|_{u\rightarrow0},
\end{equation}
as proposed in~\cite{Son:2002sd}, we can obtain
\begin{equation}
\eta=\frac{L^{d}r^{d}_{+}}{16\pi G_{d+2}}.
\end{equation}
Therefore we can arrive at the famous KSS bound
\begin{equation}
\frac{\eta}{s}=\frac{1}{4\pi}.
\end{equation}

For the speed of sound, we first note that the thermodynamic
quantities of the black hole should be identified with the
quantities in the field theory side as
\begin{equation}
\{I_{E}, M, S_{BH}, T_{H}\}~~\leftrightarrow~~\{\Omega/T, E, S, T\}
\end{equation}
where $\Omega$ denotes the thermodynamic potential. Recall the
results given in Section 2,
\begin{equation}
I_{E}=-\frac{r^{z+d}_{+}L^{d}V_{d}\beta_{H}}{16\pi G_{d+2}},~~
T_{H}=\frac{(z+d)r^{z}_{+}}{4\pi},~~
M=\frac{r^{z+d}_{+}dL^{d}V_{d}}{16\pi G_{d+2}}.
\end{equation}
Then by using the fact that the thermodynamic potential $\Omega$ is
\begin{equation}
\Omega=-PV_{d},
\end{equation}
where $P$ denotes the pressure, we can obtain
\begin{equation}
P=\frac{1}{d}\frac{E}{V_{d}}=\frac{1}{d}\epsilon,
\end{equation}
where $\epsilon$ is the energy density. Thus the speed of sound is
given by
\begin{equation}
c^{2}_{s}=\frac{\partial P}{\partial\epsilon}=\frac{1}{d}.
\end{equation}
Note that in five dimensional spacetime, i.e. $d=3$, we have
$c^{2}_{s}=1/3$, which saturates the bound conjectured in
~\cite{Hohler:2009tv},~\cite{Cherman:2009tw}.

%%%%%%%%%%%%%%%%%%%%%%%%%%%%%%%%%%%%%%%%%%%%%%%%%%%%%%%%%%%%%%%%%%%%%%%%%%%%%%%%%%%%%%%%%%%%%%%%%%%%
\section{Summary and Discussion}
%%%%%%%%%%%%%%%%%%%%%%%%%%%%%%%%%%%%%%%%%%%%%%%%%%%%%%%%%%%%%%%%%%%%%%%%%%%%%%%%%%%%%%%%%%%%%%%%%%%%
There has been enormous progress on applying the AdS/CFT
correspondence, or the more general gauge/gravity correspondence to
systems in condensed matter physics. In this note we discuss several
aspects of the exact black hole solutions in asymptotically Lifshitz
spacetime. We firstly rewrite the solution proposed
in~\cite{Taylor:2008tg} in a more convenient way. Then we show that
the tidal forces in the near horizon region tend to be infinity in
the near-extremal limit, in which sense the black hole is ``naked''.
We also evaluate the Wilson loops both analytically and numerically
in the extremal and finite temperature cases. Finally, we
investigate the hydrodynamic properties of the black holes and find
that the shear viscosity and the speed of sound both saturate the
conjectured bounds.

There are several directions which worth further studying. Firstly,
the embedding of the original Lifshitz background~(\ref{1eq1})into
string theory is still unknown. However, some Lifshitz backgrounds
with different scaling behavior have been realized in string theory
in~\cite{Azeyanagi:2009pr}, where the configurations were comprised
by D3-D7 and D4-D6 branes. It may be expected that we can
embed~(\ref{1eq1}) into string theory by superposing more different
types of D-branes.

Secondly, it has been observed in~\cite{Ghaemi:2004} that the
Lifshitz fixed point has ultralocal correlators at finite
temperature. Thus it would be interesting to calculation the
correlation functions in the black hole background, following the
prescription in~\cite{Son:2002sd}, and compare the results with
those obtained in the field theory side. Furthermore, it is
necessary to build up a systematic holographic renormalization
method~\cite{Skenderis:2008dg} in such asymptotically Lifshitz
background.

Finally, an interesting model of quantum gravity was proposed by
Horava quite recently~\cite{Horava:2008jf}. In $3+1$ dimensions,
this theory has a $z=3$ fixed point in the UV and flows to a $z=1$
fixed point in the IR, which is just the classical Einstein-Hilbert
gravity theory. Furthermore, it has been found that there exist
black hole solutions in Horava-Lifshitz gravity~\cite{horavabh}.
Thus it is interesting to study the relations between the
asymototically Lifshitz black hole and black holes in
Horava-Lifshitz gravity.

\bigskip \goodbreak \centerline{\bf Acknowledgements}
\noindent We would like to thank Rong-Gen Cai, Li-Ming Cao, Qing-Guo
Huang, Yunseok Seo and Wei-shui Xu for useful discussions and kind
help. This work was supported by the Korea Science and Engineering
Foundation(KOSEF) grant funded by the Korea government(MEST) through
the Center for Quantum Spacetime(CQUeST) of Sogang University with
grant number R11-2005-021.

%%%%%%%%%%%%%%%%%%%%%%%%%%%%%%%%%%%%%%%%%%%%%%%%%%%%%%%%%%%%%
%\appendix
%%%%%%%%%%%%%%%%%%%%%%%%%%%%%%%%%%%%%%%%%%%%%%%%%%%%%%%%%%%%

%%%%%%%%%%%%%%%%%%%%%%%%%%%%%%%%%%%%%%%%%%%%%%%%%%%%%%%%%%%%%%%%%%%%%%%%%%%%%%%%

\end{document}